\documentclass[twocolumn]{aastex631}

\usepackage{todonotes}

\makeatletter
\journalinfo{Submitted to ApJ; 15 pages; 12 figures}
\makeatother

\begin{document}

\title{A Census of Compact Elliptical Galaxies in the Coma Cluster}

\correspondingauthor{Aleksandra V. Sharonova}
\email{asharonova@voxastro.org}
\correspondingauthor{Kirill A. Grishin}
\email{grishin@voxastro.org}

\author[0009-0007-2825-8758]{Aleksandra V. Sharonova}
\affiliation{Faculty of Space Research, M.V. Lomonosov Moscow State University, 1 Leninskiye Gory, 119991 Moscow, Russia}
\affiliation{Sternberg Astronomical Institute, M.V. Lomonosov Moscow State University, 13 Universitetsky prospect, 119992 Moscow, Russia}

\author[0000-0003-3255-7340]{Kirill A. Grishin}
\affiliation{Universite Paris Cite, CNRS, Astroparticule et Cosmologie, F-75013 Paris, France}
\affiliation{Sternberg Astronomical Institute, M.V. Lomonosov Moscow State University, 13 Universitetsky prospect, 119992 Moscow, Russia}

\author[0000-0002-7924-3253]{Igor V. Chilingarian}
\affiliation{Center for Astrophysics | Harvard \& Smithsonian, 60 Garden St., Cambridge, MA 02138, USA}
\affiliation{Sternberg Astronomical Institute, M.V. Lomonosov Moscow State University, 13 Universitetsky prospect, 119992 Moscow, Russia}

\author[0000-0001-8956-5953]{Gary A. Mamon}
\affiliation{Institut d'Astrophysique de Paris (UMR7095: CNRS \& Sorbonne Université), 98 bis Bd Arago, 75014, Paris, France}

\author[0000-0003-2352-3202]{Nelson Caldwell}
\affiliation{Center for Astrophysics | Harvard \& Smithsonian, 60 Garden St., Cambridge, MA 02138, USA}

\author[0000-0002-1311-4942]{Daniel Fabricant}
\affiliation{Center for Astrophysics | Harvard \& Smithsonian, 60 Garden St., Cambridge, MA 02138, USA}

\begin{abstract}

Compact elliptical (cE) galaxies are compact stellar systems with stellar masses of $10^8 \leq M_{*}/M_{\odot} \leq 10^{10}$ and radii typically $<$0.6 kpc.  Here we investigate the properties of 13 cE galaxies in the Coma cluster, six newly identified.  Our goal in this paper is to explore whether these cEs form directly in the cluster environment or are pre-processed in small groups before infalling.  We find that pre-processing in groups significantly contributes to the cE population in the Coma cluster. We analyze Hyper Suprime-Cam (HSC) $g$ band Coma images and validate our photometric measurements through comparison with Hubble Space Telescope (HST) data.  We also analyze spectroscopic data from the Dark Energy Spectroscopic Instrument (DESI).  We significantly expand the known cE population in the Coma cluster through joint photometric and spectroscopic selection.  We identify a subpopulation of cEs that likely formed in infalling groups, through their association with host galaxies, their positions on the caustic diagram, and their projected phase-space trajectories. We estimate that the central cE population will increase by 30\% within the next 0.4 Gyr, highlighting the important role of pre-processing in cE evolution.
\end{abstract}


\section{Introduction} 
\label{sec:intro}

Compact elliptical (cE) galaxies represent a relatively rare class of compact stellar systems in the local Universe. They have stellar masses of $10^8 \leq M_{*}/M_{\odot} \leq 10^{10}$ and with radii typically $<$0.6 kpc, high stellar densities. Studies of Messier~32 \citep{1944ApJ...100..137B, 1973ApJ...179..423F} led \cite{1980A&A....82..289B} to identify compact elliptical galaxies as a distinct class. 
\citet{2000PASA...17..227D} found even smaller unresolved ultra-compact dwarf (UCD) galaxies  with old stellar populations in the Fornax cluster. Later, cEs were discovered in the Abell 1689 \citep{2005A&A...430L..25M} and Abell 496 \citep{2007A&A...466L..21C} clusters. \citet{2009MNRAS.397.1816P} discovered and analyzed a sample of 7 cEs in the Coma cluster using photometric observations from the HST Treasury Survey and spectral data from MMT/Hectospec. \citet{2009Sci...326.1379C} identified 21 cEs in clusters using HST-WFPC2 photometry and SCORPIO 6-m BTA spectroscopy. \citet{2015Sci...348..418C} identified a sample of 195 cEs combining Sloan Digital Sky Survey (SDSS) photometric and spectroscopic data with GALaxy Evolution eXplorer (GALEX) UV data.  

Although wide-field ground-based observations have identified a significant number of cE candidates, with typical $1$ arcsec seeing corresponding to $\sim 0.5$ kpc at $100$ Mpc, space observatories are required for cE structure characterization.  Spectroscopic redshifts are necessary to precisely associate cEs with host groups or clusters.

\begin{figure*}
    \centering
    \includegraphics[height = \hsize]{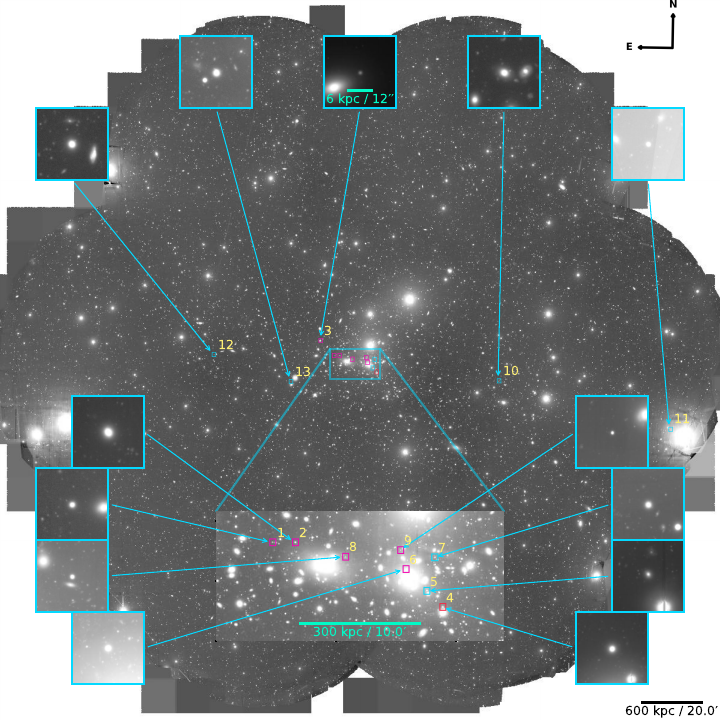}
    \caption{The position of cE galaxies from our sample in the Coma cluster. Galaxies from \citet{2009MNRAS.397.1816P} are marked as pink squares and one galaxy from \citet{2009Sci...326.1379C} and \citet{2009MNRAS.396.1647C} is marked as red square.
    }
    \label{fig:coma_map_image}
\end{figure*}

A combination of photometric and spectroscopic surveys has allowed us to place cEs in the context of compact stellar systems (CSSs), a category that also includes UCDs and globular clusters.
Most cEs are intermediate in the scaling relations (size--luminosity ($R_e - L$), velocity dispersion--luminosity ($\sigma_* - L$), and metallicity--luminosity ($[\rm{Fe/H}] - L$) between UCDs and more massive early-type galaxies \citep[ETGs; ][]{2008MNRAS.385L..83C, AIMSS_I, AIMSS_II, AIMSS_III}. cEs and UCDs may form similarly through tidal stripping.  UCDs are tidally stripped dwarf or intermediate-mass nucleated galaxies and contain nuclear star clusters surrounded by small stellar envelopes that survived stripping. Tidally stripped dwarf galaxies lack sufficient central stellar density to become cEs.  Rather, cEs are the stripped cores of elliptical \citep{1973ApJ...179..423F} or compact disc galaxy bulges \citep{2001ApJ...552L.105B, 2002ApJ...568L..13G, 2003Natur.423..519D}. Some recently formed cEs show tidal tails, clear signatures of their host galaxy interactions \citep{2011MNRAS.414.3557H, 2014ApJ...796L..14P, 2015Sci...348..418C}.  A compact stellar system inherits some properties of its tidally stripped progenitor (a larger and more massive galaxy) including central black holes more massive than expected for their stellar mass (for UCDs: \citet{2014Natur.513..398S, 2017ApJ...839...72A, 2018MNRAS.477.4856A, 2022ApJ...924...48P}; for cEs: \citet{2016MNRAS.460.1147B, 2016ApJ...820L..19P, 2017ApJ...850...15P}). An alternate, intrinsic formation scenario postulates that some UCDs represent the high-mass end of the globular cluster luminosity function \citep{2009ApJ...691..946M, 2011A&A...529A.138B, 2012A&A...537A...3M}, and that cEs are the lowest mass ETGs~\citep{1984ApJ...282...85W, 1987A&A...186...30N, 2009ApJS..182..216K, 2012ApJS..198....2K} formed via gradual build-up of stellar mass~\citep{2017MNRAS.470.4015M, 2023MNRAS.525.1192D} and mergers of nuclear star clusters~\citep{2019MNRAS.489.2746U}.

UCD and cE stellar populations provide additional clues about their evolutionary pathways. UCDs with higher metallicities likely formed through tidal stripping of intermediate-mass galaxies, whereas UCDs with lower metallicity (consistent with those observed in globular clusters) have intrinsic origins  \citep{2008MNRAS.390..906C, 2011MNRAS.412.1627C, 2012MNRAS.425..325F}. cEs forming by tidal stripping exhibit high metallicities for their mass, old stellar ages, and are often outliers in the $[\rm{Fe/H}] - L$ relation \citep{2010MNRAS.405L..11C, AIMSS_III, 2021MNRAS.503.5455F}.  Intrinsically formed cEs, on the other hand, fall on the $[\rm Fe/H] - L$ relation for ETGs \citep{2023MNRAS.525.1192D}.

A substantial fraction of the observed cEs are found in clusters \citep{2015Sci...348..418C}, but probably not all cluster cEs formed by tidal stripping in the cluster/cD potential; some galaxies could be processed before infall ~\citep{2004PASJ...56...29F}.  Some studies \citep{2009MNRAS.400..937M, 2013ApJ...770...62D, 2018ApJ...866...78H} suggest that close to half the cluster members were accreted as members of galaxy groups. 

In this work, we study the origin of the cE population in the Coma cluster with combined photometric and spectroscopic datasets. The photometric datasets consist of deep archival $g$ band Hyper Suprime-Cam \citep[HSC; ][]{2018PASJ...70S...1M} images obtained in sub-arcsec seeing.  Deep $F475W$ and $F814W$ HST/ACS Coma treasury survey images \citep{2011MNRAS.411.2439H} partially overlap the HSC fields.  The spectroscopic datasets include DESI Early Data Release~\citep{2022AJ....164..207D} and MMT/Binospec spectra.

We identify a cE subpopulation that may have been pre-processed in groups before entering the cluster environment, now contributing to the cE population in the cluster center. 

Throughout, we assume that the distance to the Coma cluster is $101\, \rm Mpc$~\citep{2003MNRAS.343..401L}, the angular scale -- $0.49\, \rm kpc\,{arcsec}^{-1}$, $H_0 = 70\,\mathrm{km\,s^{-1}}\mathrm{Mpc^{-1}}$, over density relative to critical, $\Delta_{\rm c} = 102$, the Coma cluster center is coincident with NGC4874 on the plane of the sky, with a velocity of $7000\ \mathrm{km\,s^{-1}}$ \citep{2003MNRAS.343..401L}. All magnitudes are in the AB system~\citep{1983ApJ...266..713O}, and all uncertainties are quoted at the 1$\sigma$ level unless stated otherwise.

\section{Observational Data} \label{sec:data}

\subsection{Hyper Suprime-Cam dataset} \label{sec:hsc}
Our cE selection is based on photometry from an archival HSC $g$ band Coma cluster dataset. The dataset contains 175 frames with 240~s exposures and 198 frames with 60~s exposures mostly obtained in 2016 and 2017. The images extend to radius $1.7\, \mathrm{deg}$ ($2.96\, \mathrm{Mpc}$), or 1.23$R_{\rm vir}$~\citep{2022NatAs...6..936H}, and cover $13.4 \,\mathrm{deg}^2$ with a $3\,\sigma$ magnitude limit of $25.6$. The point spread function (PSF) ranges from $0.59\, \mathrm{arcsec}$ to $1.27\, \mathrm{arcsec}$ FWHM, with a mean of $\mathrm{0.80}\, \mathrm{arcsec}$. 

The archival HSC frames were reduced with the hscPipe pipeline\footnote{\url{https://hsc.mtk.nao.ac.jp/pipedoc/pipedoc_8_e/index.html}}~\citep{2018PASJ...70S...5B}.  This pipeline returns a fully calibrated dataset with an astrometric solution and a sky background map. However, the pipeline sky background subtraction leads to dark artifacts around large galaxies, close groups and interacting intermediate-size galaxies, and over subtracts intra-cluster light. These artifacts bias the measured structural properties of the cluster members.  We built a better model of the sky background using observations of fields free of bright stars or very extended galaxies collected during the same observing runs~\citep{2023arXiv230608048C}. We will provide details of our data reduction techniques in a dedicated publication.

\subsection{DESI EDR spectra for Coma cluster} \label{sec:desi_spec}
The DESI survey observed the Coma cluster with its multi-object spectrograph \citep{2022AJ....164..207D} during the Survey Validation Program 3 (SV 3). These data were published in the Early Data Release~\citep[EDR,][]{2024AJ....168...58D}. The mean fiber allocation fraction was $99\,\%$ for the Bright Galaxy Survey~\citep[BGS,][]{2023AJ....165..253H} during SV3~\citep{2023AJ....165...50M}. This survey includes three Coma cluster components: (i)~a complete magnitude-limited sample of galaxies (BGS Bright) with $r\, <\, 19.5$; (ii)~a sample (BGS Faint) of galaxies with $19.5\,< \, r\,< \,20.175$~mag; and (iii)~a low-z AGN sample (a small contributor to the Coma galaxy sample). The BGS Bright sample probes the high-luminosity end of the cE population because it is complete to $M_{\rm r}\,<\,-15.5$~mag. The BGS Faint sample reaches $75-85\,\%$ completeness for $-15.5\,<\,M_{\rm r}\,<\,-14.8$~mag  \citep{2023AJ....165..253H} at the Coma cluster redshift.

Within the redshift range of $0.01\,<\,z\,<\,0.037$, extending to $\gtrsim 4\sigma$ of the Coma cluster's velocity dispersion ($\sigma \sim 1000\   \mathrm{km\,s^{-1}}$,~\citealp{2017ApJS..229...20S}), the DESI EDR contains $3292$ spectra.

\subsection{MMT Binospec spectra} \label{sec:binospec}

Several Coma cluster fields were observed using the Binospec spectrograph at the 6.5~m MMT in multi-object slitmask mode (mask IDs: \textit{Coma1, Coma3, ComaA, ComaB, Coma\_dE2a, Coma\_dE2b, Coma\_dE3a, Coma\_dE3b}) in 2017--2022. These datasets were obtained as part of the observing programs SAO-6, SAO-9-20a, SAO-7-21b, SAO-9-22A.  These observations were reduced with the Binospec pipeline~\citep{2019PASP..131g5005K, 2019ASPC..523..629C} in a special mode that masks extended galaxies that would contaminate the night sky model. The pipeline produces sky-subtracted, rectified, and wavelength-resampled 2D long-slit spectra calibrated in absolute fluxes, plus the corresponding flux uncertainties.  We use the Binospec spectra to validate the DESI EDR velocity dispersions and stellar population estimates, and for observations of two sample galaxies obtained with masks \textit{ComaA}\footnote{Described in \citet{2021NatAs...5.1308G}} and \textit{Coma\_dE2a}.
\textit{Coma\_dE2a} was observed on May 26 and 28, 2019 and May 20, 2020.  We used the $1000\, \mathrm{g\, mm^{-1}}$ grating, with spectral range $3805\,$\AA \ to $5310\,$\AA \ and a mean spectral resolving power of $R=5000$. The 2019 and 2020 exposure times were $2\,\rm h$ per mask, with exposure times of $4\, \rm h$ for coadded frames.

\section{Data Analysis} \label{sec:analysis}

\subsection{Automated surface photometry with SourceXtractor++} \label{sec:sepp}

We used the {\sc SourceXtractor++} package \citep{2020ASPC..527..461B, 2022arXiv221202428K} to obtain the photometric properties of the galaxies in the reduced HSC images. The code detects objects in the image and calculates initial parameters guesses and then fits a predefined combination of light profile components for all detected objects. We specify parameters for object extraction in the object detection configuration file.  Object extraction including background subtraction, filtering, segmentation, and object detection proceeds as in the earlier the {\sc SourceExtractor} package \citep{SExtractor_ref}. We set the detection threshold to $1.5$ standard deviations above the background and the minimum detection area to $5$ pixels.

The photometric modeling configuration file is written in {\sc Python}. It includes a set of photometric components, free ranges of their parameters, limits on the light fraction between the different components, a list of fixed parameters, and a list of parameters to be included to the output catalog. The parameter ranges also depend on initial guess values from the object extraction. 

We use a combination of single S\'ersic and PSF models motivated by previous studies of Coma cluster dwarf galaxies showing that the galaxy shapes cannot fit with a single S\'ersic component~\citep{2015MNRAS.453.3729H}.
For the S\'ersic profile we used the following parameters ranges: effective radius from $0.4\,\mathrm{v}$ to $10\,\mathrm{v}$ (where $\mathrm{v}$ is the initial guess) with a geometric grid; a S\'ersic index from $0.5$ to $5$ (linear steps); and an axis ratio from $0.1$ to $1$ (geometric steps).  We created the PSF frame using the {\sc PSFEx} package \citep{2011ASPC..442..435B} and the data reduction pipeline provided an error frame for the weight image.

\subsection{Analysis of DESI and Binospec spectra} \label{sec:desi}

We used the {\sc NBursts}~\citep{NBURSTs} package to perform full spectral fitting in pixel space for a sample of DESI EDR spectra. Each spectrum was modeled in the wavelength range of 3796--9800\AA~ using X-Shooter simple stellar population (SSP) templates~\citep{2022A&A...661A..50V}.

The stellar population models include a multiplicative 5th order Legendre polynomial term to fit the continuum. We account for potential emission lines by including lines with the line-of-sight velocity distribution (LOSVD) in addition to the stellar population models. In {\sc NBursts} the models are convolved with the line spread function (LSF) profile. We took the LSF profiles from the \textit{B\_RESOLUTION}, \textit{R\_RESOLUTION}, and \textit{Z\_RESOLUTION} HDUs from the {\sc coadd-*.fits} files provided by the DESI survey approximated by Gauss-Hermite (GH) polynomials. The DESI spectral resolution is nearly constant with $\sigma \approx 0.4\,$\AA, corresponding to $70\,\mathrm{km\,s^{-1}}$ at $4000$\AA~ and $30\,\mathrm{km\,s^{-1}}$ at $9500\,$\AA.

We also model Binospec 2D slit spectra with the {\sc NBursts} package. We applied adaptive binning along the slit to achieve a minimum signal-to-noise ratio in the inner and outer parts of each galaxy (see the complete description of the Binospec analysis in \citealp{2021NatAs...5.1308G}). We use the PEGASE-HR~\citep{2004A&A...425..881L} set of SSP templates, a 15th-order polynomial continuum, and a flat (zero-order) additive continuum. Our model is convolved with an LSF derived from fitting the twilight sky flat frames with a high-resolution Solar spectrum.

\begin{figure}
    \centering
    \includegraphics[width=\hsize]{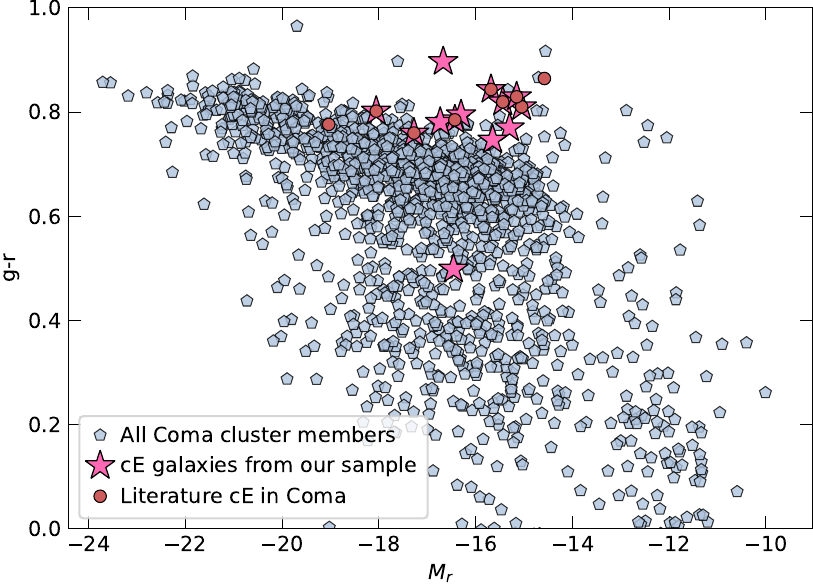}
    \caption{Color-magnitude diagram for all Coma cluster members from DESI DR1, including cE galaxies described in the literature and galaxies from our sample.}
    \label{fig:g_r_M_r}
\end{figure}

\section{Results} \label{sec:res}
\subsection{Search criteria of cEs and statistical properties of selected sample} \label{sec:ce_search}

To select a clean sample of cE galaxies we cross-matched a catalog generated by SourceXtractor++ with photometric catalog of DeCALS (Legacy Survey DR10; \citealp{2019AJ....157..168D}), spectroscopic redshift catalogs of DESI \citep{2024AJ....167...62D, 2024AJ....168...58D}, SDSS, Hectospec (as a part of RCSEDv2\footnote{\url{https://rcsed2.voxastro.org/}}), and a full spectroscopic sample of galaxies observed with the Binospec spectrograph (Sec.~\ref{sec:binospec}). The Coma cluster sample of cEs in the literature includes 9 objects, 7 galaxies from \citet{2009MNRAS.397.1816P}, and two from \citet{2009Sci...326.1379C}.

\subsubsection{Photometric selection}
\label{size_mu_crit}
Initially, we applied just the size ($R_{\rm e} < 1.2 \ \mathrm{arcsec} = 0.6$~kpc) and mean effective surface brightness ($\left\langle\mu_{\mathrm{e,g}}\right\rangle<22$~mag~arcsec$^{-2}$) criteria to our catalog of photometric properties to demonstrate that we can reproduce the sample of Coma cluster cEs from the literature. However, this selection produced a sample heavily contaminated by normal elliptical galaxies at higher redshift with colors indistinguishable from cE with red colors arising from higher metallicity. 

\begin{figure}
    \centering
    \includegraphics[width=\hsize]{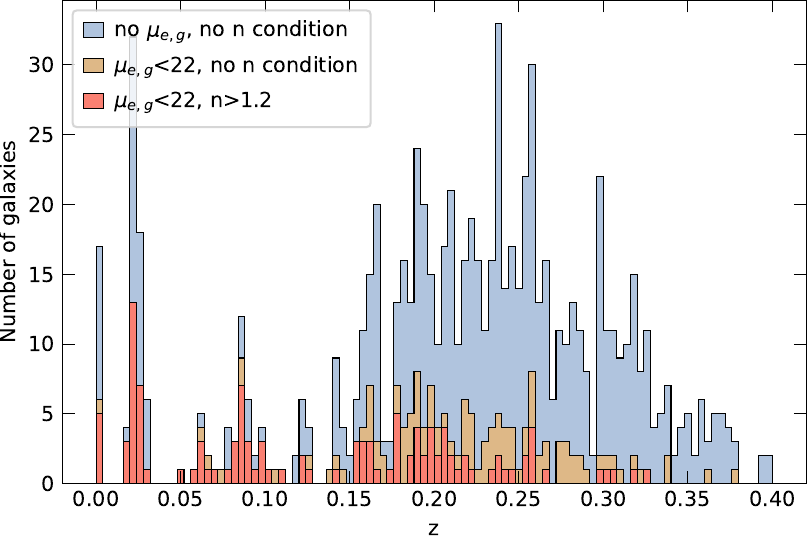}
    \caption{Distribution of redshifts for galaxies that satisfy initial criteria plus $\mu_{\rm e,g}$ and $n$ constraints.}
    \label{fig:redshift_gal_dist}
\end{figure}

We moved to a two-step selection procedure. First, we applied the magnitude ($M_r>-19.5$~mag), shape (ellipticity $(1-b/a)<0.25$), and color ($0.75<g-r<0.9$~mag) selection to the entire SourceXtractor++ catalog cross-matched against integrated photometric measurements from DESI Legacy Surveys catalog \citep{2019AJ....157..168D}. In this first step, we do not apply the surface brightness criterion. 

Selection by these three photometric criteria is $89\,\%$ complete for cE in the literature, yielding $1806$ objects, of which $916$ have redshifts from DESI or RCSED (SDSS and Hectospec). However, only $22$ of the selected galaxies are Coma cluster members. The low $2\,\%$ purity of the sample, defined as the fraction of objects that pass the selection criteria and have redshifts consistent with the Coma cluster among all objects with redshifts, is due to contamination from background star-forming galaxies at higher redshifts.

The second step in the photometric selection procedure is to reintroduce the mean effective surface brightness constraint cEs ($\left\langle\mu_{\mathrm{e,G}}\right\rangle<22$~mag~arcsec$^{-2}$) and the limit on the S\'ersic index of the outer component of the light profile ($n>1.2$). These constraints reduced the number of background objects passing the $(g-r)$ color cut criteria (Fig.~\ref{fig:redshift_gal_dist}). The completeness and purity of the sample are now $78\,\%$ and $10\,\%$ respectively.  Of $210$ objects and $132$ objects with redshifts, only 13 are Coma cluster members. The purity still remains too low to rely solely on photometric criteria to identify cE galaxies in the Coma cluster.

\subsubsection{Spectroscopic selection of the Coma cluster members}

To increase the purity of our sample, we use spectroscopic redshift measurements ($0.01\,<\,z\,<\,0.037$) and galaxy properties derived from full spectral fitting of the DESI EDR spectra. cEs typically are not forming stars and lack emission lines in their spectra. We add requirement H$\alpha$ non-detection ($F_{\rm{H}_{\alpha}} < 5 \Delta F_{H_{\alpha}}$) to the photometric size and mean effective surface brightness criteria (Section~\ref{size_mu_crit}). With the DESI spectral signal-to-noise ratio per pixel, ${\rm SNR}_{\rm spec}>2$ this limit is sensitive to star formation activity in last 50 Myr.  We did not apply color or S\'ersic index cuts because the emission line filter is more effective.

Our final sample, consisting of 13 Coma cluster cEs is presented in Table~\ref{tab:main_properties_of_cE_sample}. 
Of these, 7 were  previously known, 6 identified by \citet{2009MNRAS.397.1816P} and 1 by \citet{2009Sci...326.1379C}.

Two galaxies, GMP3511 and GMP3424, were observed with Binospec in the slit masks Coma\_dE2a and ComaA. These spectra yield better velocity dispersion and stellar population measurements than the DESI EDR data because they have much higher signal-to-noise ratio and provide spatially resolved information along the slit. 
\subsection{Photometric properties of cE galaxies}
\begin{figure}
    \centering
    \includegraphics[width=\hsize]{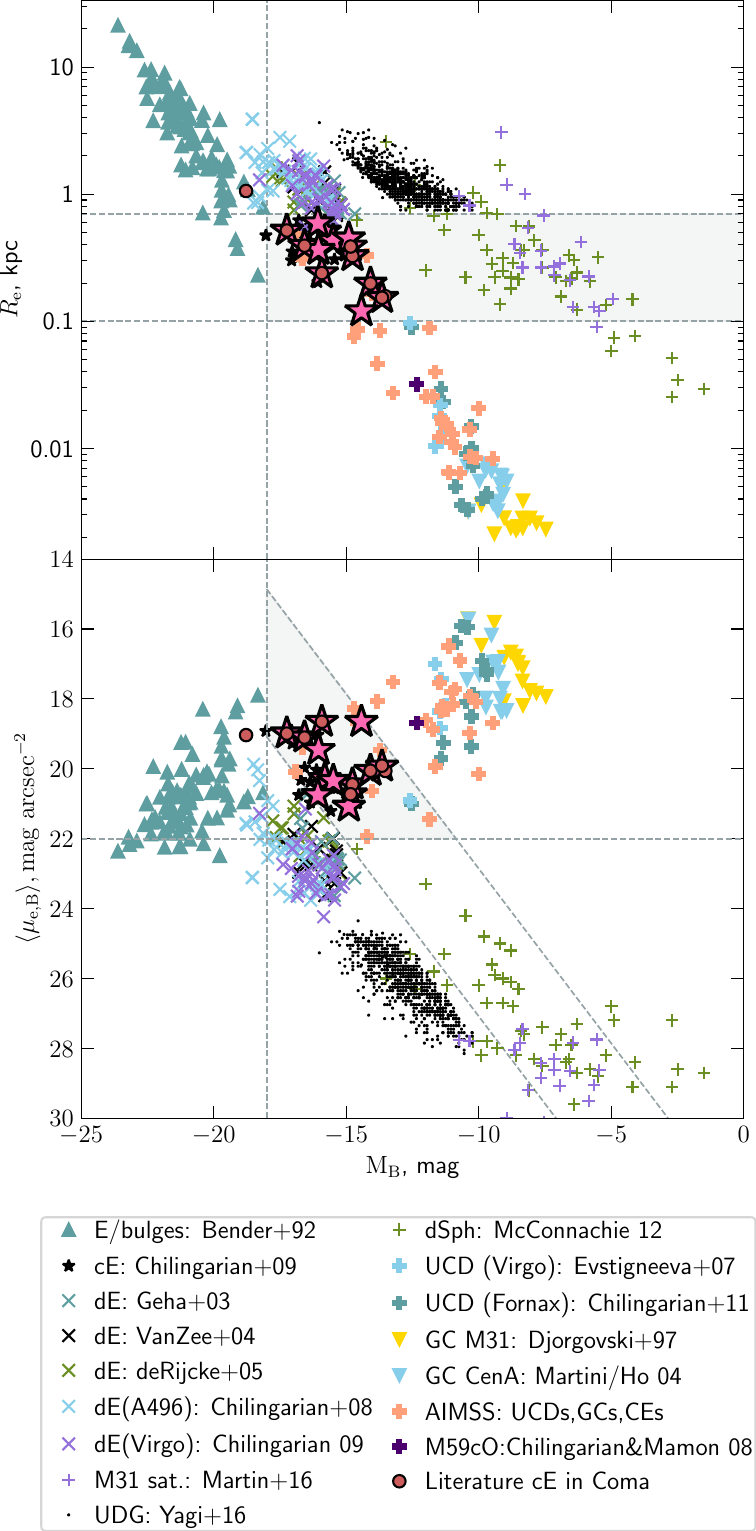}
    \caption{The size--luminosity \citep{1977ApJ...218..333K} and the surface brightness -- luminosity relations of early-type galaxies and compact stellar systems. cEs from our sample are shown by pink stars. The sources of other galaxies are shown in the legend below the plots. The grey filled areas and dashed lines denote the parameter space from Sec.~\ref{sec:ce_search}.
    \nocite{1992ApJ...399..462B} \nocite{2003AJ....126.1794G} \nocite{2005A&A...438..491D} \nocite{Chilingarian+08} 
    \nocite{Chilingarian09} \nocite{2016ApJ...833..167M}
    \nocite{yagi16} \nocite{2012AJ....144....4M}
    \nocite{2007AJ....133.1722E} \nocite{Chilingarian+11}
    \nocite{1997ApJ...474L..19D} \nocite{2004ApJ...610..233M}
    \nocite{2014MNRAS.443.1151N} \nocite{2008MNRAS.385L..83C}
    }
    \label{fig:cE_M_mueff_reff}
\end{figure}

We present the photometric properties of our 13 cEs In Table~\ref{tab:main_properties_of_cE_sample}  along with information about their hosts.  We show the position of our galaxies in the Coma cluster in Fig.~\ref{fig:coma_map_image}. Eight cEs are in the central region of the Coma cluster, while 5 are located $d_{\rm proj} > 400$  kpc from the cluster center. We place our cE galaxies in structural correlations compiled from the literature in Fig.~\ref{fig:cE_M_mueff_reff}. All lie in the cE loci in both diagrams, with an extension towards UCDs.

\begin{deluxetable*}{l|l|c|c|c|c|c|l|c|c|c}
\tablecaption{Photometric properties of the selected cEs and information about the host association. Col. 1: Galaxy identifier. Col. 2: Object name from prioritized list: 1. GMP catalog; 2. \citet{2009MNRAS.397.1816P} and 3. IAU name. Col. 3-4: Coordinates of the galaxies. Col. 5-6: Absolute magnitude and mean surface brightness within the effective radius, effective radius in the G band. Col. 7: Identified host of the galaxy (for details on the identification technique, see Section \ref{subsec:cE_infalling_groups}). Col. 8: Projected distance of the cE from the host galaxy. Col. 9: Difference between line of sight velocity of cE and potential host. Col.~10:  Projected distance from the cluster center \label{tab:main_properties_of_cE_sample}}
\tablewidth{0pt}
\tablehead{
    \colhead{\#} & \colhead{cE Name} & \colhead{RA} & \colhead{Dec} & \colhead{$M_{\rm g}$} & \colhead{$\left\langle\mu_{\mathrm{g}}\right\rangle_e$} & \colhead{$R_{\rm e}$} & \colhead{Host} & \colhead{$d_{\mathrm{proj}}$} & \colhead{$\Delta v$} & \colhead{$d_{\mathrm{projCC}}$} \\
    \colhead{} & \colhead{} & \colhead{J2000} & \colhead{J2000} & \colhead{mag} & \colhead{$\mathrm{mag\,arcsec^{-2}}$} & \colhead{kpc} & \colhead{} & \colhead{kpc} & \colhead{$\mathrm{km\,s^{-1}}$} & \colhead{kpc}
}
\startdata
1  & GMP2663              & $195.11394$ & $28.00929$ & $-14.833\pm0.004$ & $21.633\pm0.012$ & $0.388\pm0.002$ & NGC4889     & $128.62$ & $195.40$  & $347.59$ \\
2  & GMP2777              & $195.07869$ & $28.00932$ & $-17.241\pm0.001$ & $19.854\pm0.004$ & $0.518\pm0.001$ & NGC4889     & $84.16$  & $286.10$  & $294.76$ \\
3  & CcGV1                & $195.19862$ & $28.09277$ & $-14.763\pm0.010$ & $21.326\pm0.041$ & $0.326\pm0.006$ & IC4045      & $7.30$   & $193.63$  & $524.48$ \\
4  & GMP3511              & $194.84753$ & $27.91959$ & $-16.572\pm0.012$ & $19.926\pm0.040$ & $0.393\pm0.007$ & NGC4869     & $14.01$  & $228.52$  & $106.45$ \\
5  & GMP3424\tablenotemark{a} & $194.87220$ & $27.94189$ & $-15.807\pm0.006$ & $21.247\pm0.029$ & $0.508\pm0.007$ & NGC4874     & $47.93$  & $1494.84$ & $51.59$  \\
   &                      &             &            &                   &                  &                 & NGC4889     & $228.32$ & $850.33$  &         \\
6  & GMP3308              & $194.90498$ & $27.97221$ & $-15.920\pm0.008$ & $19.499\pm0.035$ & $0.239\pm0.004$ & NGC4874     & $26.62$  & $426.63$  & $24.97$  \\
7  & GMP3474              & $194.85990$ & $27.98843$ & $-15.488\pm0.018$ & $21.204\pm0.065$ & $0.430\pm0.012$ & NGC4874     & $85.98$  & $565.10$  & $79.72$  \\
8  & GMP3027              & $194.99960$ & $27.98940$ & $-14.084\pm0.008$ & $20.927\pm0.018$ & $0.198\pm0.001$ & NGC4889     & $53.95$  & $61.88$   & $166.37$ \\
9  & CcGV19b              & $194.91344$ & $27.99852$ & $-13.650\pm0.075$ & $20.800\pm0.196$ & $0.153\pm0.013$ & NGC4874     & $75.33$  & $110.93$  & $73.20$  \\
10 & GMP5500              & $194.07849$ & $27.86737$ & $-16.069\pm0.038$ & $21.311\pm0.118$ & $0.591\pm0.030$ &            &         &         & $1292.51$\\
11 & cEJ1252+2735         & $193.00710$ & $27.58707$ & $-14.426\pm0.026$ & $19.489\pm0.074$ & $0.120\pm0.004$ & PGC43618    & $34.36$  & $119.90$  & $3033.23$\\
12 & cEJ1303+2800\tablenotemark{b}  & $195.87065$ & $28.01165$ & $-16.044\pm0.002$ & $20.296\pm0.010$ & $0.366\pm0.002$ &  &  &  & $1521.22$\\
13 & GMP1993              & $195.38512$ & $27.86547$ & $-14.902\pm0.012$ & $21.886\pm0.040$ & $0.450\pm0.008$ & NGC4921     & $44.23$  & $163.01$  & $778.17$ \\
\enddata

\tablecomments{\tablenotetext{a}{For GMP3424 we provide information about possible association with both Coma cluster cDs.}
\tablenotetext{b}{cEJ1303+2800 does not have a potential host with $\Delta v < 300~\rm{km\ s^{-1}}$ within surrounding $d_{proj}=300~\rm{kpc}$. The closest massive galaxy, PGC093700, located 26 kpc away has $\Delta v = 776~\rm{km\ s^{-1}}$.}
}
\end{deluxetable*}

\subsection{cE stellar populations}

We present the cE stellar population properties derived from full DESI EDR spectrum fits in Table~\ref{tab:physical_properties_of_cE_sample}. Eight objects have velocity dispersions typical of cE galaxies, while the other 5 have lower values, possibly because they have lower stellar mass. Six galaxies have stellar ages from 7--13 Gyr and 6 have younger ages, 4--7 Gyr, indicating more recent truncation of star formation. All measured metallicities exceed $[Z/H] > -0.15$ dex. 

GMP~5500 has a measured age of $\sim$1.3~Gyr and its position on the $(g-r)$ -- $M_r$ diagram places it in the `green valley' (Fig.~\ref{fig:g_r_M_r}). GMP~5500's colors resemble those of more extended diffuse post-starburst galaxies in the Coma cluster~\citep{2021NatAs...5.1308G} and its spectrum, which cannot be described with either a simple stellar population or a truncated star-formation history requires a recent starburst. We exclude this galaxy from the discussion in the next sections but leave it on our final cE list. 

\begin{figure}
    \centering
    \includegraphics[width = \hsize]{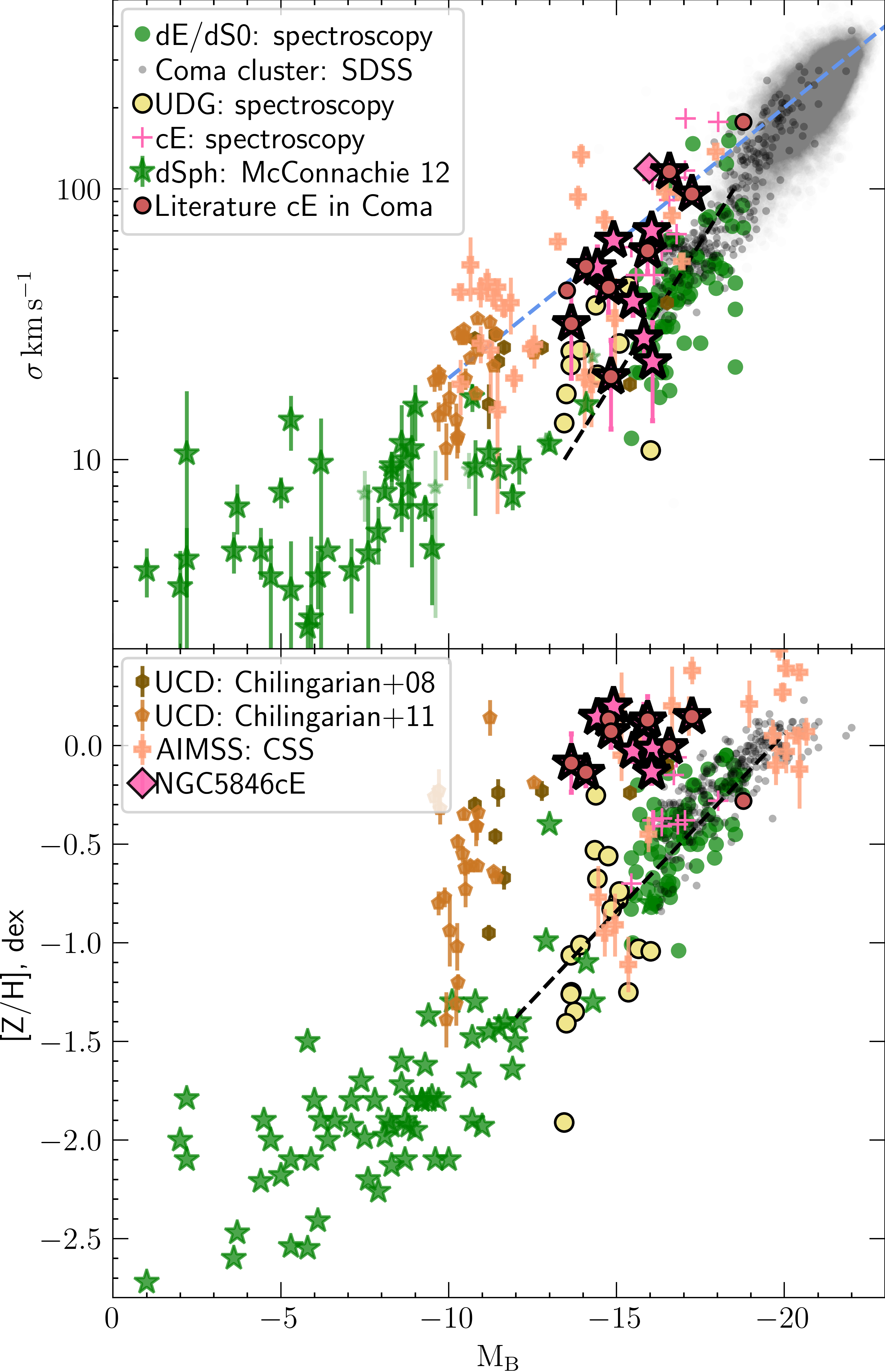}
    \caption{Relations between stellar velocity dispersion ($\sigma$, top panel), metallicity ($[Z/H]$, bottom panel) and absolute magnitude in B band ($M_B$) of early-type galaxies and compact stellar systems. Compact ellipticals (cE) from our sample are shown as pink stars. The sources of other galaxy types are indicated in the legend.}
    \label{fig:FJ_MM}
\end{figure}

\begin{deluxetable}{c|c|c|c|c}
\tablecaption{Kinematics and stellar populations of the cE sample. The properties were obtained from DESI EDR spectra  \label{tab:physical_properties_of_cE_sample}} 
\tablehead{
    \colhead{\#} & \colhead{V} & \colhead{$\sigma$} & \colhead{$t_{\rm SSP}$} & \colhead{[Z/H]} \\
    & \colhead{($\mathrm{km\,s^{-1}}$)} & \colhead{($\mathrm{km\,s^{-1}}$)} & \colhead{(Gyr)} & \colhead{(dex)}
}
\startdata
1  & $6380.4\pm4.9$   & $\ 28.2\pm7.4$   & $13.09\pm0.88$ & $\ \ 0.11\pm0.10$ \\
2  & $6217.3\pm2.6$   & $\ 95.9\pm2.5$   & $\ 5.30\pm0.16$  & $\ \ 0.15\pm0.00$ \\
3  & $6765.5\pm3.8$   & $\ 20.3\pm7.3$   & $10.46\pm0.61$ & $\ \ 0.07\pm0.08$ \\
4  & $6945.7\pm3.9$   & $116.0\pm3.9$  & $\ 7.20\pm0.11$  & $-0.01\pm0.03$ \\
5  & $5650.1\pm3.2$   & $\ 38.5\pm4.2$   & $\ 4.21\pm0.19$  & $-0.03\pm0.05$ \\
6  & $7621.5\pm3.4$   & $\ 64.7\pm3.7$   & $\ 8.07\pm0.06$  & $\ \ 0.20\pm0.00$ \\
7  & $7746.0\pm5.0$   & $\ 51.7\pm5.9$   & $\ 4.45\pm0.21$  & $-0.14\pm0.07$ \\
8  & $6585.1\pm8.3$   & $\ \ 31.8\pm11.8$  & $\ 4.97\pm0.70$  & $-0.09\pm0.15$ \\
9  & $7059.5\pm8.6$   & $\ 59.3\pm9.6$   & $\ 7.50\pm0.50$  & $\ \ 0.13\pm0.12$ \\
10 & $7233.0\pm6.1$   & $\ 43.4\pm8.3$   & $\ 1.32\pm0.01$  & $\ \ 0.13\pm0.05$ \\
11 & $8141.3\pm5.3$   & $\ 23.0\pm9.1$   & $\ 4.23\pm0.37$  & $-0.02\pm0.11$ \\
12 & $7180.9\pm4.3$   & $\ 70.3\pm4.6$   & $\ 9.34\pm0.40$  & $-0.13\pm0.06$ \\
13 & $5651.2\pm8.6$   & $\ 51.2\pm9.9$   & $\ 5.31\pm0.95$  & $\ \ 0.14\pm0.03$ \\
\enddata
\end{deluxetable}

\begin{figure}
    \centering
    \includegraphics[trim={0.1cm 0 0.0cm 0.1cm},clip,width=0.49\hsize]{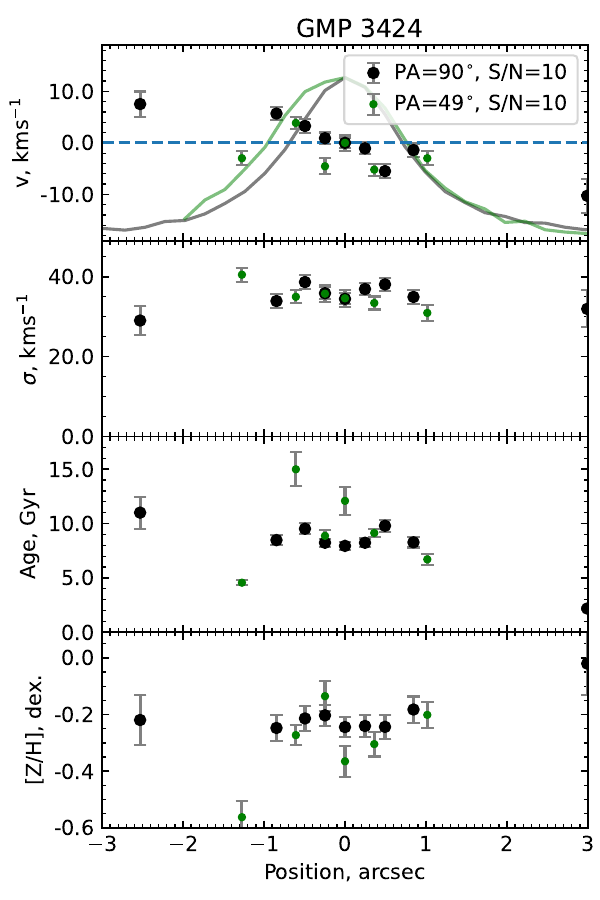}
    \includegraphics[trim={0.1cm 0 0.0cm 0.1cm},clip,width=0.49\hsize]{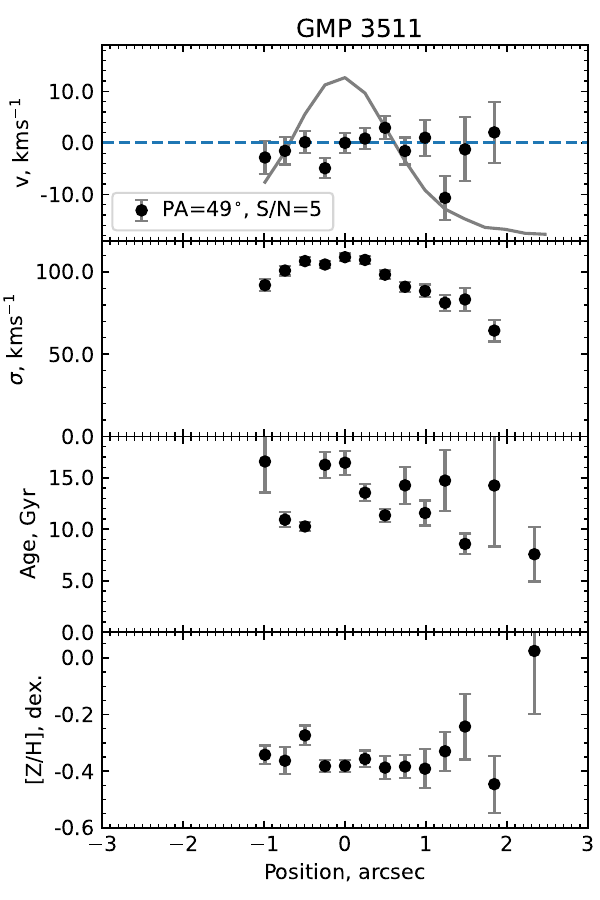}
    \caption{Profiles of the velocity ($v$), velocity dispersion ($\sigma$), age and metallicity for GMP 3424 and GMP 3511 inferred with PEGASE-HR SSP models for masks 2D spectra in Coma\_dE2a and ComaA slit masks (black and green points respectively). Legend insets contain information about the position angle (PA) of the slit and S/N. Panels with velocity ($v$) profiles (top) also show bell-shaped profiles of the mean flux extracted from the slit (grey and green lines).
    }
    \label{fig:bino}
\end{figure}

Our analysis of the GMP3424 velocity profile derived from Binospec spectra (Fig.~\ref{fig:bino})  indicates rotation with a semi-amplitude of 7 $\mathrm{km\,s^{-1}}$. The dispersion profile for this galaxy is almost flat around 35 $\mathrm{km\,s^{-1}}$ with a central dip of a few $\mathrm{km\,s^{-1}}$. The age profile shows a similar central dip, decreasing from 9.0 Gyr to 7.5 Gyr. In contrast, GMP3511 does not show any rotation and its central velocity dispersion reaches 110 $\mathrm{km\,s^{-1}}$.

\section{Discussion} \label{sec:dis}

\subsection{cE formation in infalling groups} \label{subsec:cE_infalling_groups}

We identified potential cE host galaxies for our sample, combining spectroscopic and luminosity-based criteria. For each identified cE we searched for galaxies within $300$~kpc with luminosity $M_g<-19$~mag and a velocity differing by $<$$300$ $\mathrm{km\,s^{-1}}$.  This velocity difference is the typical velocity dispersion of low-mass galaxy groups of $M_{200} \sim 10^{13} M_{\odot}$~\citep{2013MNRAS.430.2638M}. We validated the list of potential host candidates by visual inspection.  We confirmed that all of our cE galaxies with old stellar populations, except one, have hosts. The hosts are both near the center of the Coma cluster and more distant. One cE, cEJ1303+2800, is located only 26~kpc away from a more massive galaxy PGC~93700, but has a velocity difference of 776~km~s$^{-1}$ inconsistent with a gravitationally bound system.  This may be the remnant of a galaxy group destroyed by tidal interactions~\citep{2023MNRAS.518.1316H}. 

\begin{figure}
    \centering
    \includegraphics[width = \hsize]{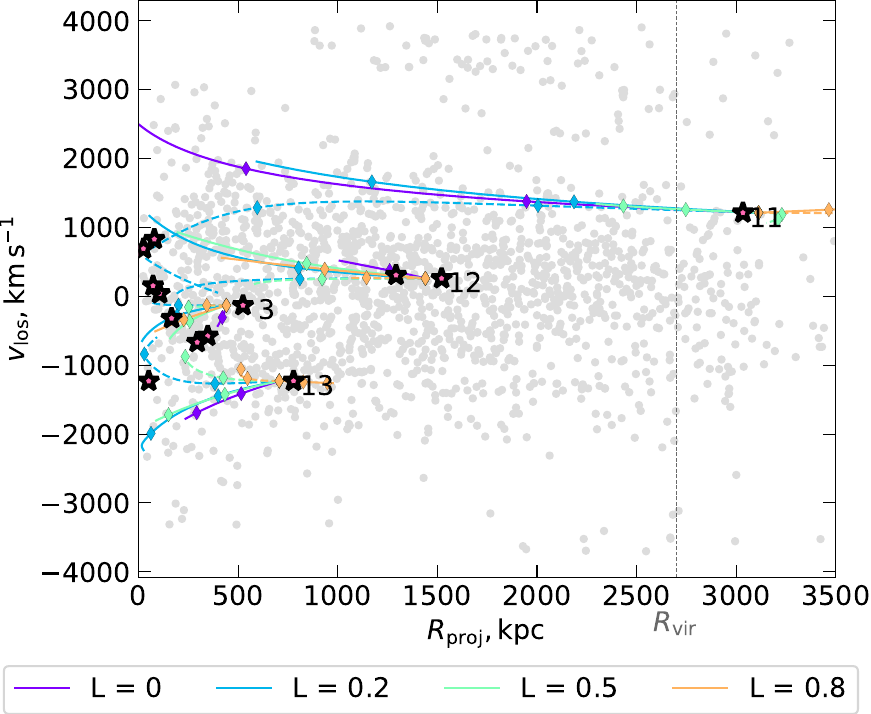}
        \caption{The caustic diagram of the Coma cluster members, along with position of all cE galaxies from our sample (pink stars) and the selected infalling cE galaxies(additionally marked by numbers according to Table~\ref{tab:main_properties_of_cE_sample}), shows possible trajectories for varying angular momentum (L). Each trajectory has timestamps. For galaxy CcGV1 (\# 3) , the timestamp is $75\,\mathrm{Myr}$; for galaxy cEJ1252+2735 (\# 11), it is $500\, \mathrm{Myr}$; and for galaxies cEJ1303+2800 (\# 12) and GMP1993 (\# 13), it is $200\,\mathrm{Myr}$. Solid lines represent trajectories at the maximum possible distance, while dotted lines correspond to the minimum possible (projected) distance for a given $L$.}
    \label{fig:caustic_diagram}
\end{figure}

We identified 4 cEs outside the central region of the cluster ($d_{proj} > 150~kpc$). Because they have massive hosts these may be cEs pre-processed in infalling groups. However, the HSC-$g$ dataset does not show any evidence for tidal interactions, including tidal tails, down to 29~mag~arcsec$^{-2}$. The absence of tidal tails suggests that putative stripping events must have occurred more than 2 Gyr ago based on simulations of the persistence of tidal tails~\citep{2019A&A...632A.122M}. In Fig.~\ref{fig:caustic_diagram} we show these cEs in the Coma cluster's caustic diagram. Two galaxies from our sample, cEJ1252+2735 and GMP1993, are located in the typically infalling portion of the caustic diagram, as indicated by a high line-of-sight velocity for their projected radius \citep{2015ApJ...806..101H}. The other two other galaxies have lower line-of-sight velocities and projected radii, and their orbital history based on the caustic diagram is ambiguous. 
These could be either infalling, with their true distances from the cluster core masked by projection, or back-splash galaxies.

\begin{figure}
    \centering
    \includegraphics[width = \hsize]{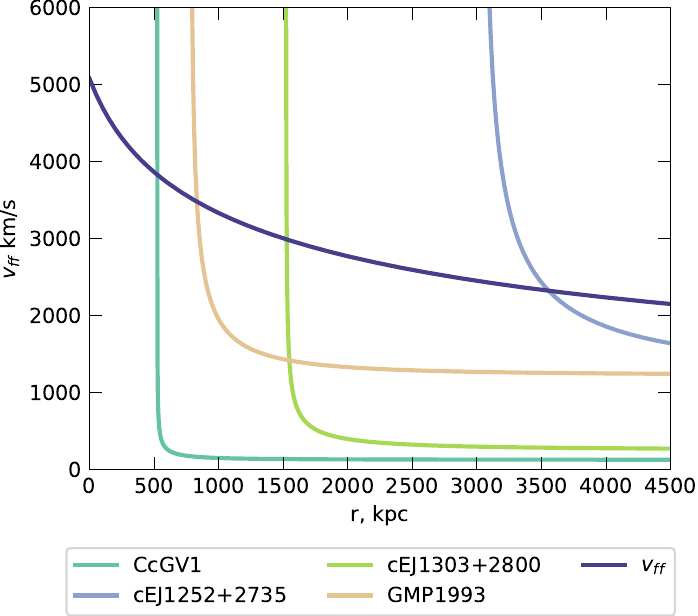}
    \caption{An illustration of the estimation of  full distances and full velocities. The purple line illustrates the free-fall velocity ($v_{ff}$) profile at different full distances from the cluster center ($r$), other lines represent possible full distances and velocities according to projected velocity and distance of the galaxies for zero angular momentum case.}
    \label{fig:explanation_of_calculation_3D_values}
\end{figure}

To estimate the potential contribution of the pre-processing channel we estimate the free-fall time of the four outer cEs assuming zero orbital energy and zero orbital angular momentum, i.e. purely radial parabolic orbits.  This estimate requires knowledge of the 3D velocity vectors and 3D distances from the center of the cluster.  The full free-fall velocities are $v_{\rm ff} (r) = \sqrt{-2\Phi(r)}$, where $\Phi$ is the gravitational potential. The total mass profile of the Coma cluster is close to that of the dark matter mass profile \citep{2003MNRAS.343..401L} because the hot inter-cluster medium and stars, contribute less than 10\% of the mass at all radii. Thus, we can adopt the analytical potential for a NFW profile:
$$
\Phi(r) = -\frac{GM_{\rm vir}}{r} \frac{\ln \left(1+\frac{r}{r_{\rm s}}\right)}{\ln(1+c)-\frac{c}{1+c}}
$$
where $r_{\rm s} = r_{\rm vir} / c$ is the scale radius, $c = 9.4$ is the concentration parameter, $r_{\rm vir} = 2.7\, \mathrm{Mpc}$, $M_{\rm vir} = 1.2\cdot10^{15}\, M_{\odot}$. We adopt these parameter values as the best-fit for the NFW profile in Table 1 of \citet{2003MNRAS.343..401L}. We show the free-fall velocity profile in Fig.~\ref{fig:explanation_of_calculation_3D_values}.  

We estimate the magnitudes of the full velocity ($v_{\rm 3D}$) and distance to the cluster center ($r_{\rm 3D}$)  for a set of possible values of angle between the plane of the sky and velocity vector ($\alpha$).   The 3D quantities are related to the observed quantities as: $$v_{\rm 3D} = \frac{v_{\rm los}}{\cos\alpha} , \ r_{\rm 3D} = \frac{r_{\rm proj}}{\sin\alpha}$$ where $v_{\rm los}$ is the line-of-sight velocity (with respect to the cluster center), $r_{\rm proj}$ is the projected distance to the cluster center. The value of $\alpha$ is determined by 3D phase space values ($r, v_{\rm 3D}$) at the intersection.

We estimate the free-fall time from energy conservation assuming zero orbital energy and negligible angular momentum:
$$t_{\rm ff} = \int_{0}^{r_0} \frac{dr}{\sqrt{v_0^2 +2\left(\Phi(r_0)-\Phi(r)\right)}}$$ 
where $r_0$ is the initial three-dimensional distance and $v_0$ is the initial 3D velocity calculated as described above. 
We present the calculated free-fall times, free-fall velocities, and 3D positions in Table~\ref{tab:galaxy_properties_ff}.

\begin{deluxetable}{llccccc} 
\tablewidth{0pt}
\tablecaption{Projected and full 3D velocities ($v_{\rm LOS}$ and $v_{\rm 3D}$ respectively), distances ($R_{\rm proj}$ and $R_{\rm 3D}$ respectively) and the free-fall time ($t_{ff}$) under assumption of infalling orbits. The free-fall time is estimated assuming zero orbital energy and zero angular momentum \label{tab:galaxy_properties_ff}}
\setlength{\tabcolsep}{2pt}
\tablehead{
\colhead{\#} & \colhead{Name} & \colhead{$v_{\rm LOS}$} & \colhead{$R_{\rm proj}$} & \colhead{$v_{\rm 3D}$} & \colhead{$R_{\rm 3D}$} & \colhead{$t_{\rm ff}$} \\
 & & \colhead{($\mathrm{km\,s^{-1}}$)} & \colhead{(kpc)} & \colhead{(km s$^{-1})$} & \colhead{(kpc)} & \colhead{(Myr)}
}
\startdata
3 & CcGV1 & -131 & 525 & 3817 & 525 & 122 \\ 
11 & cEJ1252+2735 & 1213 & 3033 & 2327 & 3555 & 1204 \\ 
12 & cEJ1303+2800 & 260 & 1521 & 2992 & 1527 & 423 \\ 
13 & GMP1993 & -1228 & 778 & 3482 & 832 & 206 \\ 
\enddata
\end{deluxetable}

\citet{2011MNRAS.416.2882M} studied the dependence of galaxy properties on their velocities and positions in and near the galaxy clusters, using SDSS data and cosmological hydrodynamical simulations \citep{2004MNRAS.348.1078B}. They plotted the distribution of physical radii versus projected distances for particles in a stacked mock cluster in LOS velocity bins. 
The galaxies CcGV1 and cEJ1303+2800 have low absolute LOS velocities (within the range $[0,\sigma_{\rm vir}]$, taking $\sigma_{\rm vir}\sim 1000\, \mathrm{km\,s^{-1}}$  for the Coma cluster~\citealp{2017ApJS..229...20S}). The galaxies cEJ1252+2735 and GMP1993 have absolute LOS velocities within $1-2\,\sigma_{\rm vir}$. The projected distance of GMP1993 is $\sim 0.2 \, R_{\rm vir}$ and according to fig.~15 in \citet{2011MNRAS.416.2882M}, the 3D physical distance in such a case remains close to the projected distance. For cEJ1252+2735, located around $1\, R_{\rm vir}$, the physical and projected distances could differ significantly. Therefore, our estimated $R_{\rm 3D}$ values agree with fig.~15 in \citet{2011MNRAS.416.2882M}.

\citet{2015ApJ...806..101H} suggest that the time-scale required for galaxies to reach their first pericenter after being accreted is approximately $0.5-0.8\, \mathrm{Gyr}$ based on observations and simulations. They considered a galaxy to be accreted as it crosses $R_{200}$ on infall.  Our calculated free-fall time for galaxies within $R_{200}$ (excepting galaxy \#11) is $0.1-0.4\, \mathrm{Gyr}$. These freefall times are lower because we measured the time from the observed galaxy positions inside $R_{200}$. Our measurements are therefore consistent with the \citet{2015ApJ...806..101H} pericenter passage times.

For cEJ1252+2735, \citet{2016MNRAS.463.3083O} defined infall as the moment when a galaxy passes $2.5 R_{\rm vir}$ ($R_{\rm vir} = 2.7\, \mathrm{Mpc}$ for the Coma cluster \citealp{2003MNRAS.343..401L}). It takes $2.5-3.0\, \mathrm{Gyr}$ to reach $R_{\rm vir}$, and an additional $1.0-1.5\, \mathrm{Gyr}$ to reach the pericenter. The estimated $R_{\rm 3D}$ for cEJ1252+2735 is $3.5\,\mathrm{Mpc}$, placing it between $R_{\rm vir}$ and $2.5 R_{\rm vir}$. The estimated free-fall time is $1.2\, \mathrm{Gyr}$, including the time required to reach the virial radius and the time to then reach the pericenter. We conclude that all estimated free-fall times are consistent with simulations and observations. 

\cite{2017MNRAS.471.4170T} estimate the orbital times of galaxies from their entry into the cluster to their passage through pericenter based on cosmological simulations with cluster mass growth through filament accretion. For the most probable infall case, with apocentric distance ($R_{apo}$) much larger than a virial radius, $R_{apo} \gg R_{vir}$, the orbital times are less than 0.5 Gyr, consistent with our estimates.

We also examined possible effects of non-zero angular momentum by solving the second-order equation of motion:

$$\frac{d^2 \mathbf{r}}{dt^2} = -\frac{G M(r)}{r^3}\mathbf{r} $$

The initial state includes six parameters: $x,y,z, v_x, v_y, v_z$, where $x$ and $y$ are the coordinates in the plane of the sky and $v_z$ is the line-of-sight velocity. These parameters are obtained directly from observations. For the case of zero angular momentum, $r_{\rm 3D}$ and $v_{\rm 3D}$ can be well constrained. 

\begin{figure}
    \centering
    \includegraphics[width=\hsize]{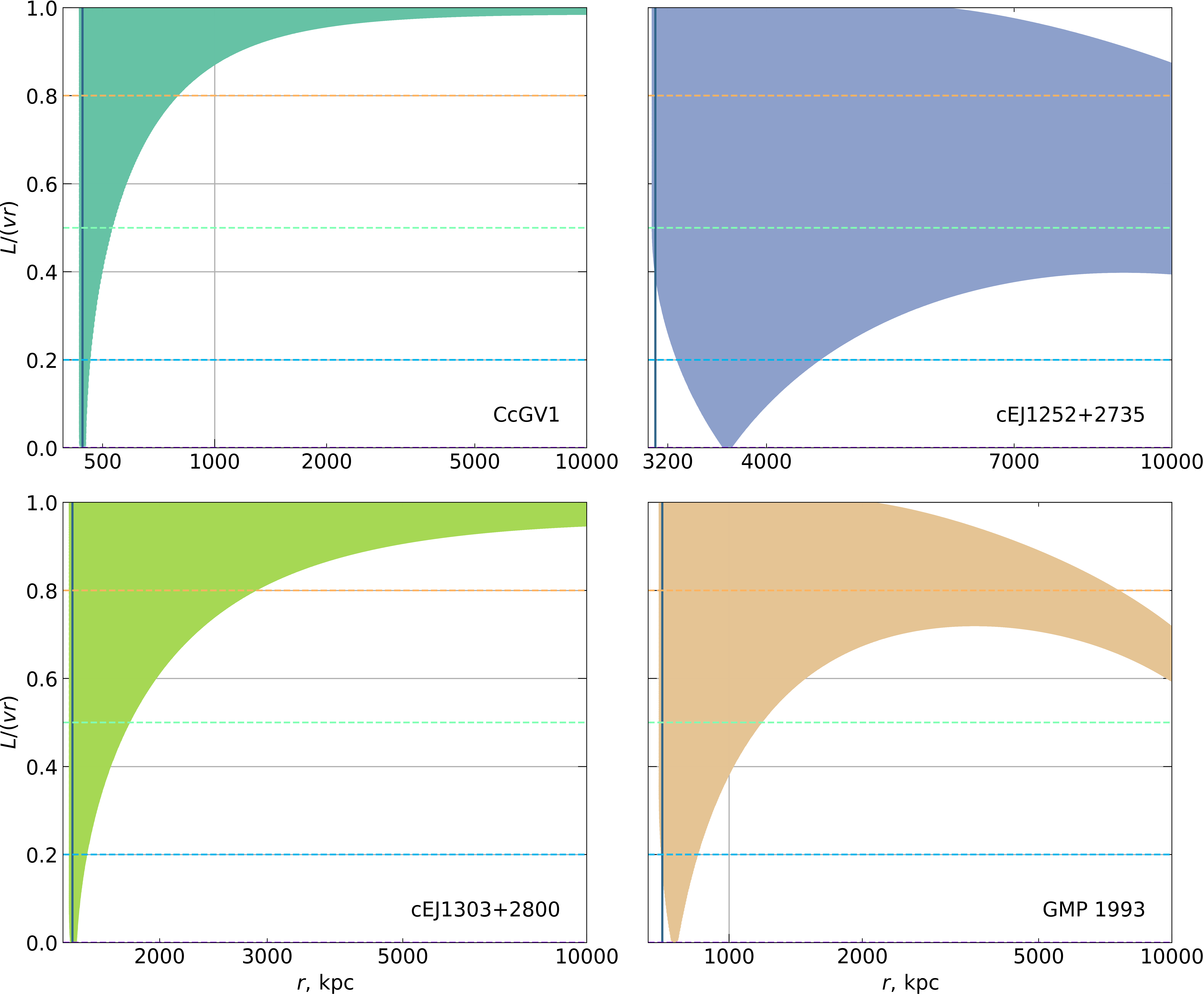}
    \caption{Possible values (filled) of dimensionless angular momentum ($L$) and total distance to the cluster center ($R$) for the four cEs in outer parts of the cluster.}
    \label{fig:calc_v_R_for_L}
\end{figure}

We define a range of possible $z$ values $-10\, \rm Mpc$ to $+10\, \rm Mpc$ and calculated $r_{\rm 3D}$ and $v_{\rm ff\, 3D}$ for non-zero angular momentum, to identify the possible parameter space for the  dimensionless angular momentum $L = {\left| \mathbf{v}_{\rm 3D} \times \mathbf{r}_{\rm 3D} \right| }/{ \left( v_{\rm 3D}\, r_{\rm 3D} \right)}$ and the full distance to the cluster center. We also compute all possible values of $v_x$ and $v_y$ for a given $v_{\rm ff\, 3D}$ and $v_z$. In Fig.~\ref{fig:calc_v_R_for_L} we plot possible values for $L$ and $R$ given observed $x$, $y$ and $v_z$.

We determine the minimal and maximal possible distances for a given normalized angular momentum value and plot possible corresponding trajectories: dotted lines represent the minimum possible distance, solid lines the maximum distance. The trajectory for each galaxy is calculated over the time needed to reach the pericenter. All outer cEs reach the inner 500~kpc for $L<$0.5.

The calculations suggest that the population of cE galaxies in the center of the Coma cluster is expected to grow by $10-30\,\%$ over the next $0.4\, \mathrm{Gyr}$.  Pre-processing in the infalling groups is therefore likely to be a major contributor to the growth of the cE population in cluster center. 

\subsection{Origin of the cE population in the Coma cluster center}

Using a source catalog based on deep HSC-$g$ images for the Coma cluster we have identified a bimodal population of cE galaxies: a dominant population of cEs in the cluster center and a subpopulation of cEs distributed in the outer parts of the cluster. All of the catalog galaxies outside the cluster center, except one, are hosted by more massive galaxies.  This observation is consistent with cE pre-processing in groups, predominantly tidal stripping.  We conclude that if cEs initially form in infalling groups, these groups will be destroyed by tidal interactions with the gravitational potential of individual galaxies and the cluster leaving the infalling cEs to contribute to the cE population in the cluster center. The cE galaxies with HST data in our sample do not host globular cluster (GC) systems. Galaxies in the stellar mass range $2\times10^8-4\times10^9$ $M_{\odot}$ are expected to host up to 50 GCs~\citep{2016ApJ...830...23B}. Their absence serves as further evidence for tidal stripping, which is expected to destroy GCs.

Our estimate suggests that pre-processing in groups will contribute approximately 20\% to the growth of the cE population over the next 0.5~Gyr. 
This estimate assumes that the cEs we observe in the outer part of the cluster have zero orbital energy and low angular momentum. Because two cE groups are located the phase-space populated mostly by the infalling galaxies, we conclude that our assumptions are plausible. Hydrodynamical simulations show that the vast majority of infalling groups do not survive their first passage in the cluster and become unbound after $0.5-1 \, \mathrm{Gyr}$ after entering the cluster~\citep{2023MNRAS.518.1316H}. Therefore, the cE hosting groups are likely to become unbound well before the accretion of cEs by their host galaxies.

\begin{acknowledgments}
AS acknowledges the Russian Science Foundation (RScF) grant No.~23-12-00146. KG acknowledges support from ANR-24-CE31-2896. IC's research is supported by the Telescope Data Center at the Smithsonian Astrophysical Observatory and by the Smithsonian Institution FY24 combined call grant.
\end{acknowledgments}

\appendix

\section{Measurements of effective radii}\label{appendix:Re_measurements}

\begin{figure}
    \centering
    \includegraphics[width=0.45\textwidth]{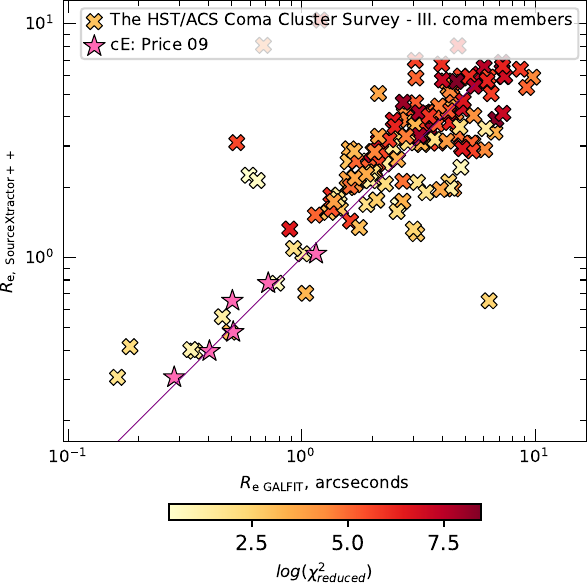}
    \includegraphics[width=0.45\textwidth]{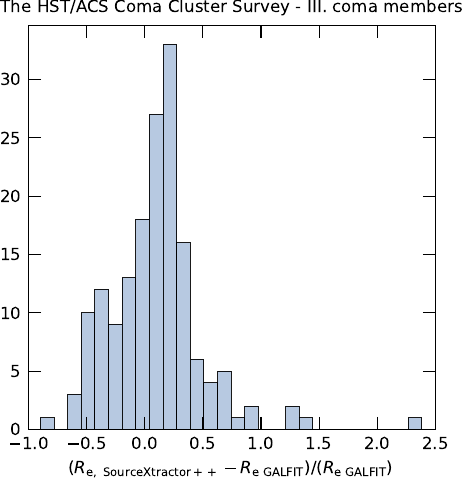}
    \caption{Measurements of $R_{\rm e} $ with SourceXtractor++ applied to HSC-g images compared to the literature {\sc Galfit} measurements on HST data from \citet{2011MNRAS.411.2439H} and \citet{2009MNRAS.397.1816P}, along with histogram of the relative differences between measurements. 
}
    \label{fig:Re_measurements}
\end{figure}

We compare our effective radius measurements with the published catalog based on the decomposition of high-resolution Hubble Space Telescope (HST) images with the {\sc Galfit}~\citep{2010AJ....139.2097P} code. For this comparison, we choose Coma cluster members with SourceXtractor++  half-light radii larger than their $3\sigma$ ncertainties ($R_{\rm e}>3\sigma_{R_{\rm e}}$). In Fig.~\ref{fig:Re_measurements} we plot the differences (${(R_{\rm e} - R_{\rm e\, lit})}/{R_{\rm e\, lit}}$) between the SourceXtractor++ and {\sc Galfit}-based $R_{\rm e}$ measurements. SourceXtractor++ $R_{\rm e}$ estimates tend to be systematically higher than {\sc Galfit}'s. 
The median relative difference (bias, $\Delta R / R$) is 0.13 with a dispersion of 1.16. A possible explanation of this discrepancy is the difference between the fitting models. We fit galaxies with two- or even three-component models, while in \citet{2011MNRAS.411.2439H} use a single S\'ersic profile. For cE galaxies, the two-component model is superior to a single-component model because most galaxies from the comparison sample contain embedded structures or nuclei. A second contributor to the discrepancy may be the use of different photometric bands. \citet{2011MNRAS.411.2439H} analyzed \textit{F814W} images ($I$ band), while we used $g$-band data: radial color gradients often observed in early-type galaxies (bluer outer regions compared to centers) will naturally explain larger $R_e$ values in a bluer band.

\begin{figure}
    \centering
    \includegraphics[width=0.45\textwidth]{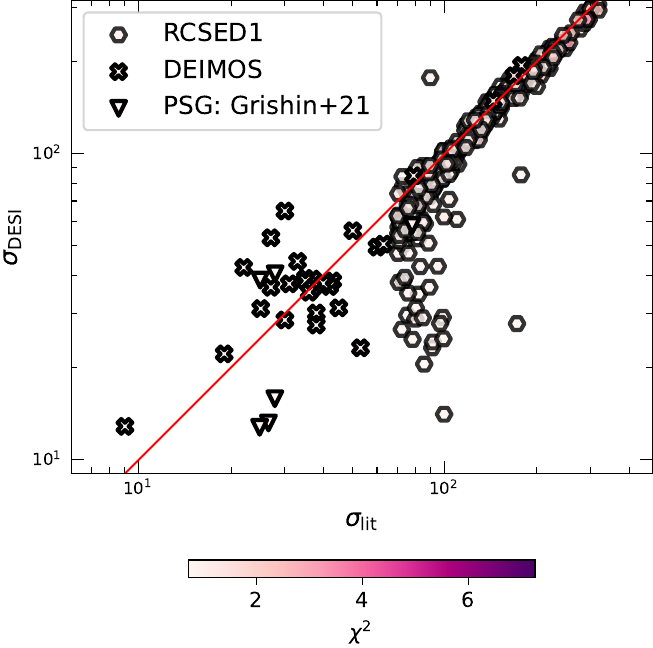}
    \includegraphics[width=0.45\textwidth]{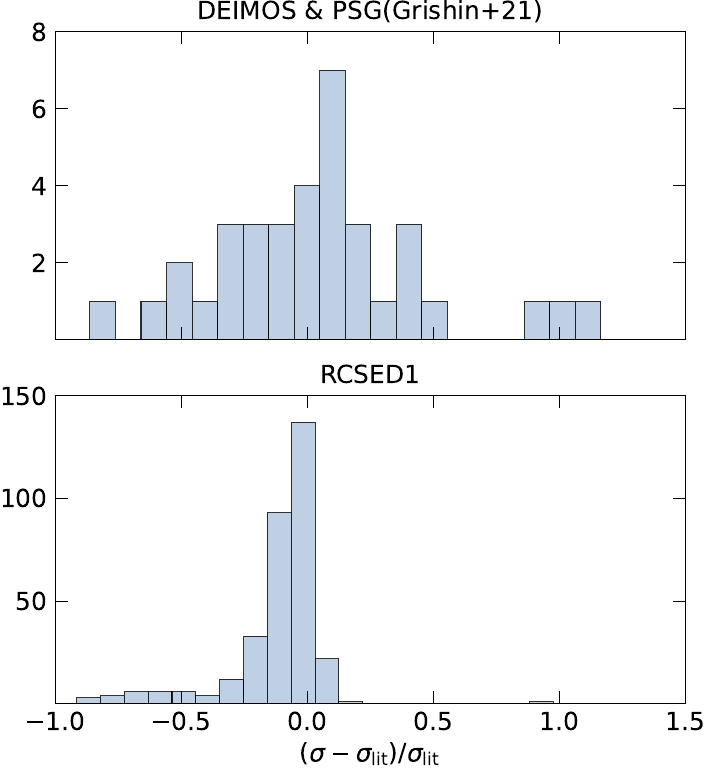}
    \caption{Velocity dispersion measurements of DESI compared to RCSED1 (SDSS sample) \citep{2017ApJS..228...14C} and DEIMOS \citep{2012MNRAS.420.2819K} measurements, along with histrogram of the relative difference.}
    \label{fig:vel_disp_comparison}
\end{figure}

\begin{figure*}
    \centering
    \includegraphics[height = \hsize]{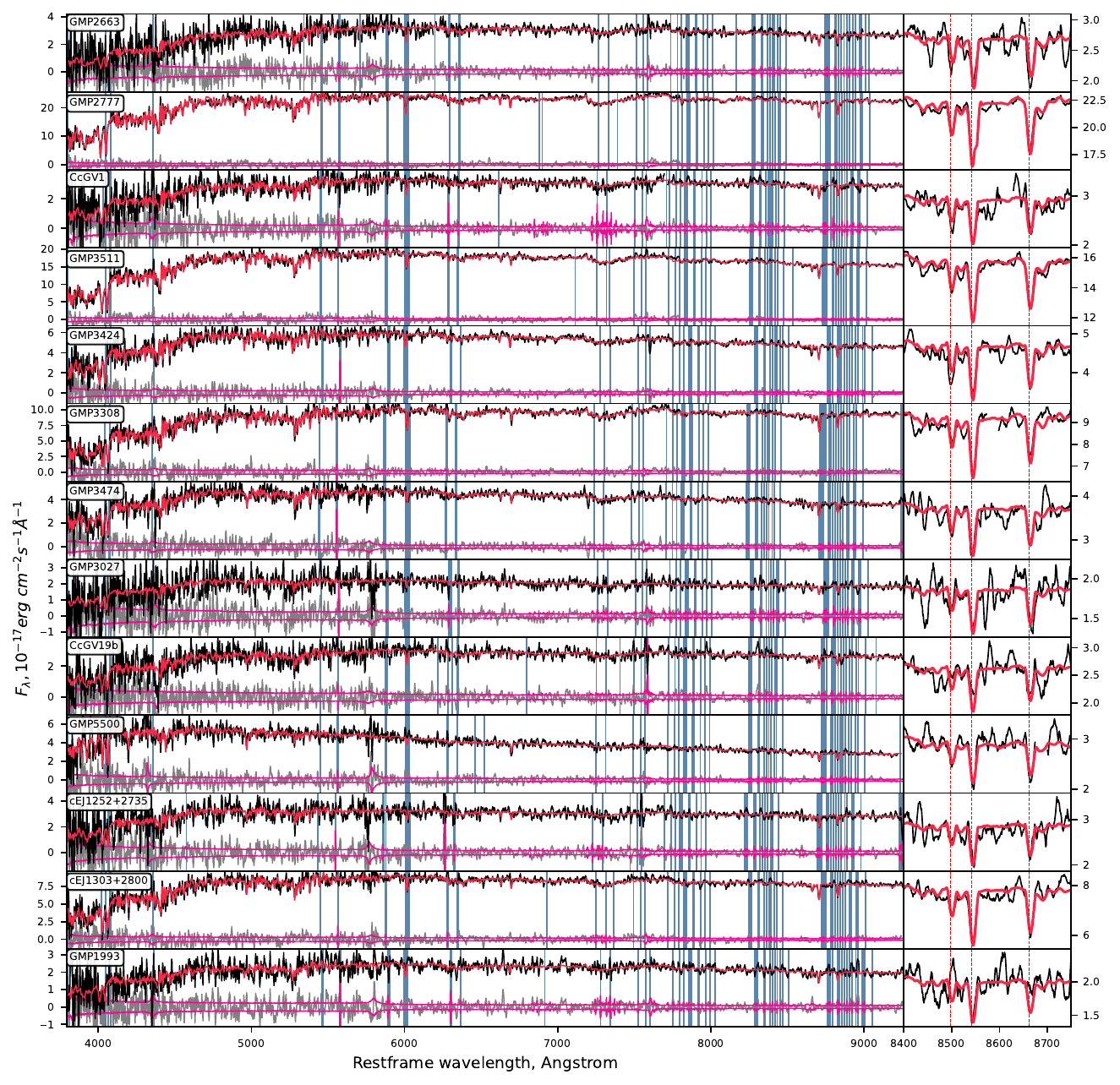}
    \caption{
    DESI spectra of cEs optimally extracted along the slit flux  calibrated in $F_{\lambda}$ units (black). Their best-fitting stellar population models (red), flux uncertainties (pink), and fitting residuals (gray). Galaxy names from Tab.~\ref{tab:main_properties_of_cE_sample} are shown in each panel. All data and best-fitting models are smoothed with the Savitzky–Golay filter for display purposes. The right panel illustrates the fit to the calcium triplet region.
    }
    \label{fig:DESI_spec_table}
\end{figure*}

\bibliography{cE_Coma_DESI}
\bibliographystyle{aasjournal}

\end{document}